\documentclass[aps,twocolumn,floatfix,showpacs,superscriptaddress]{revtex4}
\usepackage{amsmath,amsfonts,amssymb,graphicx,natbib}

\begin{document}

\title{The logarithmic negativity: A full entanglement monotone that is not convex}

\author{M.\ B.\ Plenio}

\affiliation{QOLS, Blackett Laboratory, Imperial College London,
Prince Consort Road, London SW7 2BW, UK}

\affiliation{Institute for Mathematical Sciences, Imperial College
London, 53 Exhibition Road, London SW7 2BW, UK}


\begin{abstract}
It is proven that the logarithmic negativity does not increase on
average under general positive partial transpose preserving (PPT)
operation (a set of operations that incorporate local operations
and classical communication (LOCC) as a subset) and, in the
process, a further proof is provided that the negativity does not
increase on average under the same set of operations. Given that
the logarithmic negativity is {\em not} a convex function this
result is surprising as it is generally considered that convexity
describes the local physical process of losing information. The
role of convexity and in particular its relation (or lack thereof)
to physical processes is discussed and the importance of
continuity in this context is stressed.
\end{abstract}

\date{\today }

\maketitle

{\em Introduction ---}
Entanglement is the key resource in many quantum information
processing protocols. Therefore it is of interest to develop a
detailed understanding of its properties. In view of the resource
character of entanglement it is of particular interest to be able
to quantify entanglement
%With this aim in mind, basic properties
%of so-called entanglement measures have been identified in the
%literature
\cite{Plenio V 98,Horodecki 01,Horodecki H01,Eisert P
03,Plenio V 05}.

Any resource is intimately related to a constraint which the
resource allows to overcome. Therefore, the detailed character of
entanglement and its quantification as a resource depends on the
constraints that are being imposed on the set of operations. In a
communication setting where two spatially separated parties aim to
manipulate a joint quantum state it is natural to restrict
attention to local quantum operations and classical communication
(LOCC). In this case separable states are freely available while
non-separable states, which cannot be prepared by LOCC alone,
become a resource. The phenomenon of bound entanglement
\cite{Horodecki HH 98}, i.e. non-separable states that possess a
positive partial transpose and are useless to distill pure
maximally entangled states by LOCC \cite{Peres 96,Horodecki HH
96}, suggest that there are other natural restricted classes of
operations. One might add bound entangled states as a free
resource to LOCC operations to achieve tasks that are impossible
under LOCC alone \cite{Audenaert PE 03,Ishizaka 04,Ishizaka P 05}.
Encompassing both classes is the mathematically more natural and
convenient class of positive partial transpose preserving
operations (PPT-operations) \cite{Rains 01} which have the
property that they map the set of positive partial transpose
states into itself just as LOCC operations map the set of
separable states into itself. Under this set of operations
distillable quantum states become a valuable resource while bound
entanglement is free.
%In this work we will prove any statement for both, LOCC and PPT
%operations.

\begin{figure}[th]
\centerline{
\includegraphics[width=8.2cm]{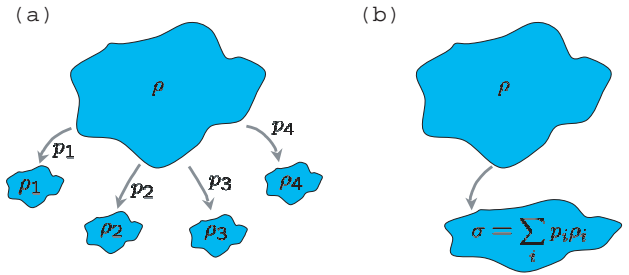}
} \vspace{-.cm} \caption{\label{subselect} Schematic picture of
the action of quantum operations with and without sub-selection
shown in part (a) and part (b) respectively.  }
\end{figure}

A function $E$ that is suggested to quantify entanglement must
satisfy certain conditions. Apart from the requirement that $E$
vanishes on the set of states that can be created using LOCC (or
PPT) alone, the most important property is that of the {\em
non-increase on average} of $E$ under LOCC (or PPT)
\cite{Horodecki 01,Horodecki H01,Eisert P 03,Plenio V 05,Plenio V
98}, ie
\begin{equation}
    E(\rho) \ge \sum_i p_i E(\rho_i)
    \label{monotonicity}
\end{equation}
where, in a LOCC (PPT) protocol applied to state $\rho$, the state
$\rho_i$ with label $i$ is obtained with probability $p_i$ (see
part (a) of fig. \ref{subselect} for an illustration). Note that
eq. (\ref{monotonicity}) is more restrictive than the requirement
that entanglement decreases under the less general set of
operations implementing $\rho\rightarrow \sigma=\sum_i p_i\rho_i$
(see part (b) of fig. \ref{subselect}), ie that $E(\rho) \ge
E(\sum_i p_i\rho_i)$. Restricting attention to such operations
would imply an additional constraint, namely that we are unable to
select sub-ensembles according to a measurement outcome. Such an
additional constraint is not directly related to the non-local
structure of quantum mechanics and would obscur key features of
entanglement. Therefore we consider here, as in the bulk of the
literature, condition eq. (\ref{monotonicity}).

A quantity $E$ possessing the above properties, in particular eq.
(\ref{monotonicity}), is called an entanglement monotone. If such
a monotone is furthermore identical to the entropy of entanglement
on pure states \cite{Bennett BPS 96}, ie the entropy of the
reduced density operator of one party, then it is called an
entanglement measure \cite{Plenio V 05}.

A significant number of such entanglement monotones and
entanglement measures have been formulated and their properties
have been explored \cite{Bennett BPS 96,Bennett DSW 96,Vedral PRK
97,Horodecki 04,Vedral P 98,Wootters 98,Plenio VP 00,Horodecki HH
00,Virmani P 00,Audenaert EJPVD 01,Hayden HT 01,Plenio V
01a,Audenaert DVW02,Donald HR 02,Tucci 02,Vidal W 02,Eisert AP
03,Christandl W 04,Horodecki HHOSSS 04}. While many of these
quantities, for example the entanglement cost and the distillable
entanglement, are operationally motivated, their mathematical
formulation generally involves an optimization over
high-dimensional spaces which makes their evaluation exceedingly
difficult. Even when the minimization involves a convex function
on a convex set \cite{Vedral PRK 97,Vedral P 98} so that the
optimization is numerically feasible, analytical expressions are
generally not available except in cases of very high symmetry
\cite{Eisert FPPW 00,Vollbrecht W 01}.

It is therefore of some interest to find quantities that are
entanglement monotones, ie satisfy eq. (\ref{monotonicity}) and
are at the same time easy to compute. Such quantities are valuable
even if for pure states they do not coincide with the entropy of
entanglement and/or lack an operational interpretation. One
example is the so-called negativity \cite{Werner 98,Zyczkowski HSL
98,Eisert P 99}, to be defined below, which has been proven to be
an entanglement monotone \cite{Lee KPL 00,Vidal W 02,Eisert
01,Plenio unpub}. Unfortunately, the negativity does not possess a
striking operational interpretation. Its close cousin the
logarithmic negativity, on the other hand, is an upper bound to
distillable entanglement and has an operational interpretation
\cite{Audenaert PE 03}. The logarithmic negativity is, however,
not convex. This implies that it can increase under mixing, a
process which is often considered to describe the loss of
classical information, ie a local process. Therefore it was
believed that the logarithmic negativity is not an entanglement
monotone in the sense of eq. (\ref{monotonicity}) (see for example
page $3$ of \cite{Vidal W 02}). However, this reasoning is not
conclusive and in fact incorrect.

Following some basic definitions, we discuss the relation of
convexity to physical processes in general and the loss of
classical information and highlight the importance of continuity
in this context. Then, we proceed to present a proof that
demonstrates that indeed both negativity {\em and} the logarithmic
negativity are entanglement monotones, satisfying eq.
(\ref{monotonicity}), both under LOCC and under PPT operations.

{\em Notations and Definitions ---} For any operator $A$ we define
the trace norm $||A||_1 = tr|A| = tr\sqrt{A^{\dagger}A}$, ie the
sum of the singular values of $A$ \cite{Bhatia 97,Horn J 87}.
Employing the trace norm we then define the {\em negativity} as
\begin{equation}
    N(\rho) = \frac{ || \rho^{\Gamma_A} ||_1-1}{2}
\end{equation}
where $\rho^{\Gamma_A}$ (in the following we will drop the index
$A$) denotes the partial transpose of $\rho$ with respect to party
A. This definition ensures that the negativity vanishes on
ppt-states and coincides with the entropy of entanglement on
maximally entangled states. Note that the negativity differs from
the entropy of entanglement for all other pure entangled states.
It is known that the negativity is an entanglement monotone under
general LOCC operations as well as PPT-operations in the sense of
eq. (\ref{monotonicity}) \cite{Lee KPL 00,Vidal W 02,Eisert
01,Plenio unpub}.

A more easily interpreted and useful quantity is obtained by
considering the {\em logarithmic negativity} \cite{Werner 98}
which is defined by
\begin{equation}
    LN(\rho) = \log ||\rho^{\Gamma}||_1 \; .
\end{equation}
This quantity exhibits monotonic behaviour under LOCC and
PPT-operations $\Psi$ in the sense $LN(\Psi(\sigma))\le
LN(\sigma)$, ie in processes not involving sub-selection. It is an
upper bound to distillable entanglement \cite{Vidal W 02} and
possesses an operational interpretation as a special type of
entanglement cost under PPT operations \cite{Audenaert PE 03}. For
general LOCC or PPT operations it is not known however whether the
logarithmic negativity is an entanglement monotone, ie whether eq.
(\ref{monotonicity}) is satisfied. Indeed, it was suggested (see
for example page $3$ of \cite{Vidal W 02}) that the lack of
convexity of the logarithmic negativity implies the existence of
LOCC operations that increase the logarithmic negativity on
average. We will now discuss why the lack of convexity alone is
not sufficient to destroy monotonicity in the sense of eq.
(\ref{monotonicity}), highlight the importance of continuity and
then present a rigorous proof for the monotonicity of both the
negativity {\em and} the logarithmic negativity.

{\em Convexity Issues ---} In the present context it is important
to note that the convexity requirement itself is not
straightforwardly connected to the physical process of discarding
of information \cite{Plenio V 05,Eisert FPPW 00}. The loss of
information refers to a situation where one begins with a
selection of locally identifiable states $\rho_i$ that appear with
rate $p_i$ to end up in a mixture of these states which is of the
form $\rho=\sum p_i \rho_i$. Indeed, the first situation, before
the loss of information about the state, can be described by the
quantum state
\begin{equation}
    \sum_i p_i |i\rangle_M\langle i|\otimes \rho_i^{AB},
\end{equation}
where $\{|i\rangle_M\}$ denote some orthonormal product basis.
Clearly a measurement of the marker particle $M$ reveals the
identity of the state of parties $A$ and $B$ without disturbing
the associated states $\rho_i$. The loss of information about the
identity of the states $\rho_i$ is then described by tracing out
the marker particle $M$ to obtain $\rho=\sum p_i \rho_i$
\cite{Plenio V 05,Plenio V 01}. This process should not increase
entanglement and we would like to see that
\begin{equation}
    E(\sum_i p_i |i\rangle_M\langle i|\otimes \rho_i^{AB}) \ge E(\rho)
\end{equation}
is satisfied. Indeed, this requirement is a special case of eq.
(\ref{monotonicity}). It is important to note however that this
process is not identical to the mathematically convenient
convexity requirement
\begin{equation}
    \sum_i p_i E(\rho_i) \ge E(\sum_i p_i \rho_i).
\end{equation}
Indeed, one can explicitly demonstrate that the logarithmic
negativity is a concave function on Werner states and can
therefore increase under mixing of quantum states \cite{Audenaert
PE 03}. One might argue however, that there is a connection
between mixing and the loss of information in the asymptotic
limit. Asymptotic mixing, ie the process
\begin{displaymath}
 \otimes_{i=1}^{k} \rho_1^{\otimes p_i N} \rightarrow
    (\sum_{i=1}^{k} p_i \rho_i)^{\otimes N}
\end{displaymath}
can be realized in the limit $N\rightarrow \infty$ with arbitrary
precision using only LOCC \cite{Vidal 02}. This appears to suggest
that mixing and the loss of information are identical but one
should note that this is only so for quantities that possess
sufficiently strong continuity properties in the asymptotic limit.
This is not obviously so for the logarithmic negativity as is
already suggested by the lockability of the logarithmic negativity
\cite{Horodecki HHO 04}. As the detailed continuity properties of
the logarithmic negativity are not known we are not able to
connect the convexity with local loss of information here.
Therefore the monotonicity of the logarithmic negativity is an
open question that needs to be settled directly.

{\em Monotonicity properties ---} We will now prove that the
logarithmic negativity as well as the negativity are entanglement
monotones in the sense of eq. (\ref{monotonicity}) both under
general LOCC operations as well as the more general
PPT-operations.
% In fact, we will obtain the stronger property of monotonicity
% under PPT operations.

{\em Lemma:} The logarithmic negativity is an entanglement
monotone, satisfying eq. (\ref{monotonicity}), for general trace
preserving completely positive PPT operations.\\
{\em Proof:} The proof proceeds in two steps. First we consider
the monotonicity properties of $||\rho^{\Gamma}||_1$. For
trace-preserving completely positive PPT operations, denoted
CP-PPT) this will exhibit the same behaviour as the negativity
itself \cite{Ishizaka P 05a}. Then we will proceed to demonstrate
that this implies the monotonicity of the logarithmic negativity.

Let us consider a completely positive PPT-operation $\Psi$ that
maps $\rho$ to $\sigma=\Psi(\rho)$ deterministically and denote
with $A_{+}$ ($A_{-}$) the positive (negative) part of the
operator $A$. Employing the linearity of $\Psi$ and the fact that
$\Psi$ maps positive states to positive states, we find
\begin{eqnarray}
    tr|\Psi(\rho)| &=& tr \{\Psi(\rho)_{+} - \Psi(\rho)_{-}\}\nonumber\\
    &=& tr \{\Psi(\rho)_{+} + \Psi(-\rho)_{+}\}\nonumber\\
%    &\le& tr \{\Psi(\rho_{+}\} + tr \{\Psi(-\rho))_{+}\}\nonumber\\
    &\le& tr \{\Psi(\rho_{+})\} + tr \{\Psi((-\rho)_{+})\}\nonumber\\
    &=& tr \{\Psi(\rho_{+})\} + tr \{\Psi(-(\rho)_{-})\}\nonumber\\
    &=& tr \{\Psi(\rho_{+}-\rho_{-})\}\nonumber\\
    &=& tr \Psi(|\rho|).\label{positive}
\end{eqnarray}
Now consider a general CP-PPT operation that maps $\rho$ to
$\rho_i=\Psi_i(\rho)/tr \Psi_i(\rho)$ with probability $p_i=tr
\Psi_i(\rho)$ such that all the $\Psi_i$ are CP-PPT maps and
$\sum_i \Psi_i$ is trace preserving. Employing eq.
(\ref{positive}) we find
\begin{eqnarray}
    \sum_i p_i ||\rho_i^{\Gamma}||_1
%    &=& \sum_i tr \Psi_i(\rho) tr|\frac{(\Psi_i(\rho))^{\Gamma}}{tr \Psi_i(\rho)}|
%    \nonumber\\[-0.1cm]
    &=& \sum_i tr|(\Psi_i(\rho))^{\Gamma}|\nonumber\\
    &=& \sum_i tr|\Psi_i^{\Gamma}(\rho^{\Gamma})|\nonumber\\
    &\le& \sum_i tr \Psi_i^{\Gamma}(|\rho^{\Gamma}|)\nonumber\\
    &=& \sum_i tr (\Psi_i(|\rho^{\Gamma}|^{\Gamma}))^{\Gamma}\nonumber\\
%    &=&  tr \sum_i\Psi_i(|\rho^{\Gamma}|^{\Gamma})\nonumber\\[-0.1cm]
%    &=& tr \sum_i \Psi_i(|\rho^{\Gamma}|^{\Gamma})\nonumber\\
    &=&  tr |\rho^{\Gamma}|^{\Gamma}\nonumber\\
    &=& ||\rho^{\Gamma}||_1 \, .\label{monotone}
\end{eqnarray}
This demonstrates the monotonicity of
$||\rho^{\Gamma}||_1=tr|\rho^{\Gamma}|$.

To prove the monotonicity, in the sense of eq.
(\ref{monotonicity}), of the logarithmic negativity we use eq.
(\ref{monotone}), the concavity of the logarithm and its
monotonicity to obtain
\begin{eqnarray*}
    \sum_i p_i LN(\rho_i) &=& \sum_i p_i \log_2 ||\rho_i^{\Gamma}||_1 \nonumber\\
    &\le& \log_2 \sum_i p_i ||\rho_i^{\Gamma}||_1\\
    &\le& \log_2 ||\rho^{\Gamma}||_1\nonumber\\
    &=& LN(\rho) \, .
\end{eqnarray*}
This is the monotonicity of the logarithmic negativity under
general PPT-operations in the sense of eq. (\ref{monotonicity})
which completes the proof.

{\em Summary and Conclusions ---} We proved that the negativity is
an entanglement monotone both under LOCC, for which we also
provide an alternative, previously unpublished proof \cite{Plenio
unpub} complementing existing proofs \cite{Lee KPL 00,Vidal W
02,Eisert 01}, and under the more general setting of positive
partial transpose preserving operation. We extended this result
further to prove that also the logarithmic negativity, which
possesses an operational interpretation \cite{Audenaert PE 03}, is
an entanglement monotone both under general LOCC and PPT
operations. This is despite the logarithmic negativity being
neither convex nor concave, a fact that has previously led to the
expectation that the logarithmic negativity cannot be a full
entanglement monotone. The key observation however is that
convexity is merely a mathematical requirement for entanglement
monotones and generally does not correspond to a physical process
describing the loss of information about a quantum system. Indeed,
it is the concavity in combination with the monotonicity of the
logarithm that permits the proof of the non-increase of the
logarithmic negativity under PPT operations.

{\it Acknowledgments ---} The author is grateful for discussions
with K. Audenaert, D. Browne, J. Eisert, S. Ishizaka, S. Virmani
and R.F. Werner on the subject of this paper. Careful reading and
encouraging comments by M. Christandl are gratefully acknowledged.
This work is part of the QIP-IRC (www.qipirc.org) supported by
EPSRC (GR/S82176/0) as well as the EU Thematic Network QUPRODIS
(IST-2001-38877) and the Leverhulme Trust (F/07 058/U).


\begin{thebibliography}{99}
\bibitem{Ishizaka P 05a} For non-trace preserving CP-PPT operations
this is however unclear although results in \cite{Ishizaka P 05}
suggest that monotonicity of the negativity might hold even under
non-trace preserving PPT operations.
\end{thebibliography}
\end{document}